\documentstyle[12pt]{article}

\begin{document}

{\bf Two approximate formulae for the binding energies in }$\Lambda -${\bf %
hypernuclei and light nuclei.}

\begin{center}
Th.E.Liolios\footnote{%
email:theoliol@physics.auth.gr}

{\footnotesize Department of Theoretical Physics,University of
Thessaloniki,Thessaloniki 54006,Greece}
\end{center}

{\bf Abstract}

Two approximate formulae are given for the binding energies in $\Lambda -$%
hypernuclei and light nuclei by means of the (reduced) Poeschl-Teller and
the Gaussian central potentials. Those easily programmable formulae combine
the eigenvalues of the transformed Jacobi eigenequation and an application
of the hypervirial theorems.

\vspace{0.5cm} {\it PACS: 21.80.+a,}\vspace{0.5cm} 21.10.Dr

A wealth of experimental results has been produced in an attempt to
determine the binding energies of a $\Lambda -$hyperon in $\Lambda -$%
hypernuclei defined as:

\begin{equation}
E=\left( M_{core}+M_{\Lambda }\right) c^{2}-M_{H}c^{2}  \label{bindener}
\end{equation}

where $M_{\Lambda },M_{core}$ and $M_{H}$ are the masses of the
hypernucleus, its core and of the $\Lambda -$hyperon respectively. Those
experiments range from the nuclear emulsions to the strangeness exchange
reactions \cite{povh,bruckner,bertini1} and up to the more recent associated
production reaction. \cite{milner,chrien1,pile1,chrien2,pile2}.

The importance of those energies stems from the fact that the $\Lambda -$%
particle, unlike other elementary particles (like pions or K-mesons), is a
probe of the inner nuclear density and not of the nuclear surface. Moreover
the binding effects of the $\Lambda -$hyperon influence the average
properties of the nucleus such as the nuclear rms radius. An accurate
theoretical determination of the $\Lambda -$binding energies is, therefore,
of paramount importance. In fact various methods have been employed using
phenomenological $\Lambda -$nucleus potentials or self-consistent
calculations. Although the derived theoretical formulae are satisfactory,
more convenient methods are needed for a handy interpretation of the
experimental results. That need becomes more intense when it comes to the
study of the mass number dependence of the energy quantities in $\Lambda -$%
hypernuclei and light nuclei.

The reduced Poeschl-Teller (RPT) central potential :

\begin{equation}  \label{rpt}
V\left( r\right) =-\frac{V_{0}}{\cosh ^{2}\left( \frac{r}{R}\right) }
\end{equation}

and the Gaussian one:

\begin{equation}
V\left( r\right) =-V_{0}e^{-\frac{r^{2}}{R^{2}}}  \label{gap}
\end{equation}
can be used as approximations to the self consistent potential in the study
of the behavior of a nucleon in light nuclei $\left( \cite{bu}\right) $and
have also been used extensively in Hypernuclear Physics $\left( \cite{lalphd}%
-\cite{cpc}\right) $. In fact the above potentials belong to the general
class :

\begin{equation}
V(r)=-V_{0}f(\frac{r}{R})\hspace{1cm}0\leq r<\infty  \label{genpot}
\end{equation}
where $V_{0}>0$ is the potential depth, $R>0$ the potential radius and the
''potential form factor'' $f$, $\left( f(0)=1\right) ,$ is an even analytic
function of $x=\frac{r}{R}.$ Namely, 
\begin{equation}
f(x)=\sum\limits_{k=0}^{\infty }d_{k}x^{2k}
\end{equation}
where $d_{k}$ are the numbers: 
\begin{equation}
d_{k}=\frac{1}{(2k)!}\frac{d^{2k}}{dx^{2k}}f(x)\mid _{x=0}\hspace{1cm}%
k=0,1,2,...\hspace{1cm}d_{1}<0
\end{equation}

However, the corresponding Schroedinger eigenvalue problems:

Poeschl-Teller:

\begin{equation}  \label{pteq}
\left[ \frac{d^{2}}{dx^{2}}-\frac{l(l+1)}{x^{2}}+s^{-2}\cosh ^{-2}\left(
x\right) +\stackrel{\sim }{E}_{nl}\right] u_{nl}=0
\end{equation}
\[
u_{nl}\left( 0\right) =0\qquad u_{nl}\left( \infty \right) =0 
\]

Gaussian:

\begin{equation}
\left[ \frac{d^{2}}{dx^{2}}-\frac{l(l+1)}{x^{2}}+s^{-2}e^{-x^{2}}+\stackrel{%
\sim }{E}_{nl}\right] u_{nl}=0
\end{equation}
\[
u_{nl}\left( 0\right) =0\qquad u_{nl}\left( \infty \right) =0 
\]

where, $x=\frac{r}{R},$ $s=\left( \frac{\hbar ^{2}}{2\mu V_{0}R^{2}}\right)
^{\frac{1}{2}}$ and $\stackrel{\sim }{E}_{nl}=s^{-2}\frac{E_{nl}}{V_{0}},$%
are not exactly solvable ( but for the $s-$states of the Poeschl-Teller one)

By means of the Hypervirial theorems(HVT), a method used extensively in
Theoretical Physics$\left( \cite{liolint}-\cite{witwit}\right) $, there were
obtained approximate analytic expressions for the eigenvalues of the
potentials in question in the form of $s-$series. For instance for the RPT
potential the eigenvalues are :

\[
\stackrel{\sim }{E}_{nl}^{HVT}=-1+2a_{nl}s-\frac{1}{12}\left[
12a_{nl}^{2}-4l(l+1)+3\right] s^{2}+\frac{a_{nl}}{60}\left[
15-4l(l+1)\right] s^{3}- 
\]
\[
-\frac{4}{945}l(l+1)\left[ 12a_{nl}^{2}-4l(l+1)+3\right] s^{4}-\frac{a_{nl}}{%
907200}\left\{ 33280a_{nl}^{2}l(l+1)-\right. 
\]
\[
\left. -12816\left[ l(l+1)\right] ^{2}+25\left[ 440l(l+1)+567\right]
\right\} s^{5}-\frac{1}{1247400}l(l+1)\left\{ 32720a_{nl}^{4}+\right. 
\]
\begin{equation}  \label{ptenl}
\left. +8a_{nl}^{2}\left[ 1445-828l(l+1)\right] +\left[ 3-4l(l+1)\right]
\left[ 1084l(l+1)-2445\right] \right\} s^{6}+...
\end{equation}

where $a_{nl}=2n+l+3/2$

Moreover, there has been pointed out that the Jacobi eigenequation (JEQ) :

\begin{equation}  \label{jeq}
\left[ \frac{d^{2}}{dx^{2}}-\frac{l(l+1)}{\sinh ^{2}\left( x\right) }+\frac{%
s^{-2}}{\cosh ^{2}\left( x\right) }+\varepsilon _{nl}^{J}\right] u_{nl}=0
\end{equation}

\[
u_{nl}\left( 0\right) =0\qquad u_{nl}\left( \infty \right) =0 
\]

with 
\begin{equation}  \label{jeigval}
\varepsilon _{nl}^J=-\left[ a_{nl}-\sqrt{\frac 1{s^2}+\frac 14}\right] ^2
\end{equation}
can be used as an approximation to the Schroedinger eigenvalue problem for
the Gaussian potential\cite{besbes}, thus obtaining very accurate
approximate analytic eigenfunctions for the $s-$ states of that potential%
\cite{jag}.

The similarity between $\left( \ref{jeq}\right) $ and $\left( \ref{pteq}%
\right) $ is obvious . In fact if it wasn't for the centrifugal term of $%
\left( \ref{pteq}\right) $ they would have been identical\cite{lal93}. Note
that, for the $s$-states ($l=0),\ $series $\left( \ref{ptenl}\right) $ is
summed up to the well known exact formula for the RPT potential\cite{lalphd}%
: 
\begin{equation}
\stackrel{\sim }{E}_{n0}=-\left[ a_{n0}-\sqrt{\frac{1}{s^{2}}+\frac{1}{4}}%
\right] ^{2}=\varepsilon _{n0}^{J}  \label{enerpt}
\end{equation}

It is therefore plausible to expand $\left( \ref{jeigval}\right) $ in a way
similar to $\left( \ref{ptenl}\right) $ , that is with respect to $s$, and
study the two expansions term by term. In fact, it turns out that if one
disregards higher terms ( whose contribution is negligible anyway,
especially for relatively heavy nuclei) the first terms of that HVT-series
can be given by a very simple, easily programmable formula of a satisfactory
accuracy.

Therefore, after some algebra, the proposed approximate formula for the
eigenvlaues of the RPT potential obtained is: 
\begin{equation}
E_{nl}^{RPT}\simeq V_{0}s^{2}\left\{ \varepsilon _{nl}^{J}+\frac{L}{3}\left[
1-\frac{1}{5}a_{nl}s-\frac{16}{105}\left( a_{nl}^{2}-L\right) s^{2}-\frac{104%
}{945}a_{nl}^{3}s^{3}\right] \right\}  \label{jacptdr}
\end{equation}

where $L=l\left( l+1\right) .$

In the same spirit one can use the JEQ in order to approximate the first
terms of the corresponding HVT series for the eigenvalues of the Gaussian
central potential. In that case the approximation is ({\it very})
satisfactory only for the ({\it ground state}) $s$-states . Namely : 
\begin{equation}
E_{n0}^{Gauss}\simeq V_{0}s^{2}\left\{ \varepsilon _{n0}^{J}+\frac{1}{4}%
\left( \frac{1}{4}+a_{n0}^{2}-\frac{11}{24}a_{n0}^{3}s-a_{n0}s-\frac{2}{9}%
a_{n0}^{4}s^{2}\right) \right\}  \label{jaga}
\end{equation}

The kinetic energy can be readily derived by applying the Hellman-Feynmann
theorem\cite{martin} which in our case takes the form:

\begin{equation}  \label{hfs}
<T>_{nl}=\frac s2\frac{\partial E_{nl}}{\partial s}
\end{equation}

To assess the accuracy of those formulae two tables have been set up which
depict the energy eigenvalues for various $\Lambda -$hypernuclei calculated
through $\left( \ref{jacptdr}\right) $ and $\left( \ref{jaga}\right) $
respectively, against those calculated

a) by a numerical integration of the corresponding Schroedinger equation.

b) by means of a perturbation method\cite{lal94}

c) experimentally\cite{mildover}

The parameters of the potentials used here were obtained by fitting
procedures to the available experimental data.

In the tables we use the following notation:

$A=$The mass number of the core nucleus of the hypernuclei

$n=$The principal quantum number

$l=$ The angular momentum quantum number

$E_{nl}^{num}=$The energy eigenvalues obtained by numerical integration

$E_{nl}^{anal}=$The energy eigenvalues obtained by the present formulae

$E_{nl}^{\exp }=$The energy eigenvalues( $\Lambda -$binding energies)
obtained experimentally.

$E_{nl}^{pert}=$The energy eigenvalues obtained through a perturbation method

Hence, for the RPT potential we obtain the following table :

\[
\stackrel{\bf TABLE\ I}{
\begin{array}{cccccccc}
& A & n & l & E_{nl}^{anal}\left( RPT\right) & E_{nl}^{num}\left( RPT\right)
& E_{nl}^{\exp } & E_{nl}^{pert}\left( RPT\right) \\ 
_{\Lambda }^{28}Si & 27 & 0 & 1 & -6.86 & -6.63 & -7.0\pm 1 & -7.02 \\ 
_{\Lambda }^{31}S & 31 & 0 & 1 & -7.76 & -7.56 & -8\pm 0.5 & -7.92 \\ 
_{\Lambda }^{51}V & 50 & 0 & 1 & -10.77 & -10.66 & -12\pm 1 & -11.05 \\ 
_{\Lambda }^{51}V & 50 & 0 & 2 & -3.94 & -3.88 & -4.0\pm 1 & -4.11 \\ 
_{\Lambda }^{89}Y & 88 & 0 & 1 & -14.07 & -14.01 & -15.5\pm 1.0 & -14.23 \\ 
_{\Lambda }^{89}Y & 88 & 0 & 2 & -7.69 & -7.64 & -9.5\pm 1 & -7.90 \\ 
& 131 & 1 & 1 & -5.77 & -5.77 &  &  \\ 
& 131 & 1 & 2 & -1.67 & -1.74 &  & 
\end{array}
} 
\]

According to Table.I, the proposed formula for the RPT potential gives
satisfactory results especially for the lower excited states of \ relatively
heavy nuclei.Note that the perturbation method gives no result for the $%
2p,2d $ states (or higher ones with nodes).

As for the Gaussian potential the corresponding table is:

\[
\stackrel{\bf TABLE\ II}{
\begin{array}{cccccccc}
& A & n & l & E_{nl}^{anal}\left( Gauss\right) & E_{nl}^{num}\left(
Gauss\right) & E_{nl}^{\exp } & E_{nl}^{pert}\left( Gauss\right) \\ 
_{\Lambda }^{16}O & 15 & 0 & 0 & -12.97 & -12.66 & -12.5\pm 0.35 & 12.39 \\ 
_{\Lambda }^{28}Si & 27 & 0 & 0 & -16.18 & -16.00 & -16.0\pm 0.28 & -16.19
\\ 
_{\Lambda }^{32}S & 31 & 0 & 0 & -16.87 & -16.71 & -17.5\pm 0.5 & -16.88 \\ 
_{\Lambda }^{51}V & 50 & 0 & 0 & -19.04 & -18.96 & -19.9\pm 1 & -19.05 \\ 
_{\Lambda }^{89}Y & 88 & 0 & 0 & -21.26 & -21.22 & -22.1\pm 1.6 & -21.27 \\ 
_{\Lambda }^{138}Ba & 137 & 1 & 0 & -11.02 & -11.36 &  & -11.05
\end{array}
} 
\]

Obviously, for the Gaussian potential, the approximation is extremely good
for the ground state while for the higher states there is room for
improvement. The validity of the proposed formulae can, of course, be
further assessed by their application to the experimental results [1-8],
just as it was done in the above tables.

A clear advantage of the present method, apart from its simplicity and its
application to states not covered by the perturbation method in ref.\cite
{lal94} (see Table I), is its $s$ and $L$ dependence which provides a
straightforward means of accuracy assessment. Namely Eq.$\left( \ref{jacptdr}%
\right) $ shows that the lower the excited state the better the
approximation , as the parameter $L$ becomes smaller .

On the other hand the $s$-dependence of the method isolates and clarifies
the effects of the width and depth of the potential as well as of the mass
number of the nucleus as shown in some recent studies\cite{liolepj}, \cite
{grypliol}. Thus, the deeper and wider the potential the smaller the
parameter $s$ and consequently the better the accuracy of the method. As for
the mass number of the nucleus , the heavier the nucleus the smaller the
parameter $s$ and inevitably the better the accuracy. Moreover, in an effort
to derive approximate wavefunctions of the potentials in question, one can
equate the energy eigenvalues instead of the rms radii \cite{jag},\cite
{grypliol} in order to use the parameter $s$ as a fitting one. That approach
is simpler but has not been tried yet.

That $s$-dependence was also shown \cite{liolepj} to enhance greatly the
study of the mass number dependence of the energy quantities through the
relation

\begin{equation}
R=r_{0}\left( A\right) A^{1/3}  \label{rad}
\end{equation}

where $R$ is the nuclear radius of a nucleus with mass number $A$, assuming
a rigid core model. Note that another significant advantage is that the
kinetic and the potential energy can be derived readily from $\left( \ref
{jacptdr}\right) $ and $\left( \ref{jaga}\right) $ by a simple
differentiation with respect to $s$ \cite{liolint}, through Eq.$\left( \ref
{hfs}\right) $.

Of course there are remarkable applications of that $s-$dependent method in
other fields such as in Atomic Physics\cite{kotsos}.

We should underline that the present formulae could be also used both in the
study of nucleon-nucleus interactions and whenever a handy fair
approximation of the eigenvalues of the potentials considered is needed.

Finally it would be very interesting, indeed, to apply the present method to
the general class of potentials $\left( \ref{genpot}\right) $ and study the
region of its validity which is expected to be very satisfactory for some
particular form factors.


\begin{thebibliography}{99}
\bibitem{povh}  B.Povh, Prog.Part.Nucl.Phys.5,1 (1978)

\bibitem{bruckner}  W.Br\"{u}ckner et al., Pys.Lett. 55B, 107 (1975); 79B,
157 (1978)

\bibitem{bertini1}  R.Bertini et al., Phys.Lett. 83B, 306 (1979)

\bibitem{milner}  C.Milner et al., Phys.Rev. Lett. 54, 1237 (1985)

\bibitem{chrien1}  R.E.Chrien, Proc.of XIth Int.Conf.on Particles and
Nuclei, Kyoto 1987 (Nucl.Phys.A 478, 705c (1988))

\bibitem{pile1}  P.H.Pile, Proc.of Int.Symp.on Hypernuclear and Low Kaon
Physics, Padua 2988 (Nuovo Cimento A 102, 413 (1989))

\bibitem{chrien2}  R.E.Chrien, ibid. Nuovo Cimento A.102, 823 (1989)

\bibitem{pile2}  P.H.Pile et al., Phys. Rev.Lett.66, 2585 (1991)

\bibitem{bu}  B.Buck,Nucl.Phys.A.275,246(1977)

\bibitem{lalphd}  (a) G. A. Lalazissis,M. E. Grypeos and S. E. Massen,Phys.
Rev. C 37,2098(1988); (b) G.A.Lalazissis,Ph.D thesis, Aristotle University
of Thessaloniki(1989)

\bibitem{glnp}  M.E.Grypeos,G.A. Lalazissis,S.E.Massen,C.P.Panos J.Physics
A,332 (1989)

\bibitem{lal94}  G.A.Lalazissis ,Phys. Rev. C 49,(1994)1412, and private
communication

\bibitem{lal93}  G. A. Lalazissis,Phys. Rev. C 48,198(1993)

\bibitem{liolint}  Th.E.Liolios and M.E.Grypeos International Journal of
Theoretical Physics 36 (1997) 2051

\bibitem{jag}  Th.E.Liolios M.E.Grypeos , J.Phys.A,30(1997)L325-L330

\bibitem{cpc}  Th.E.Liolios, Computer Physics Communications,105(1997)254

\bibitem{castrobook}  (a) G.Marc and
W.G.McMilllan,Adv.Chem.Phys.58,209(1985); (b) F.M.Fernandez and E.A.Castro,
Hypervirial Theorems ,Lecture notes in Chemistry,Vol.43(Springer-Verlag)
(1987); (c) S.M.McRae and E.R.Vrscay,J.Math.Phys.33,3004(1992)

\bibitem{swendan}  R.J.Swenson and S.H.Danforth,J.Chem.Phys.57,1734(1972)

\bibitem{killin}  (a) J.Killingbeck,Phys.Lett.65A,87(1978);(b)
J.Killingbeck,J.Phys.A 18,L1925(1985); (c) J.Killingbeck,J.Phys.A 18,
245(1985)

\bibitem{grantlai}  M.Grant and C.S.Lai,Phys.Rev.A\ 20,718(1979)

\bibitem{laigauss}  C.S.Lai,J.Phys.A 16,L181(1983)

\bibitem{}  C.S.Lai,Phys.Rev.A 23,455(1981)

\bibitem{}  C.S.Lai,Phys.Rev.A\ 26,2245(1982)

\bibitem{laillin}  C.S.Lai and H.E.Lin,J.Phys.A 15,1495(1982)

\bibitem{witwit}  (a)M.R.M Witwit,J.Phys.A 24,5291(1991);(b) M.R.M
Witwit,J.Phys.A 24,4535(1991);(c) M.R.M Witwit,J.Phys.A 24,3053(1991);(d)
M.R.M Witwit,J.Phys.A 24,3041(1991)

\bibitem{besbes}  N.Bessis,G.Bessis and B Joulakian,J.Phys.A 15,3679(1982)

\bibitem{martin}  R.A.Bertlmann and A.Martin ,Nuclear Phys.B, 168,111(1980)

\bibitem{mildover}  D.J.Millener, C.B.Dover, and A.Gal, Phys.Rev.C 38,2700
(1988)

\bibitem{liolepj}  Th.E.Liolios, Eur.Phys.J.A, in press

\bibitem{grypliol}  M.E.Grypeos, Th.E.Liolios, Phys.Lett.A. 34,252 (1999)

\bibitem{kotsos}  B.A.Kotsos, Th.E.Liolios, M.E.Grypeos,C.G.Koutroulos,
Physica B.,in press
\end{thebibliography}
\end{document}